\documentclass[fleqn, usenatbib]{mnras}
\usepackage{amssymb,amsmath}
\usepackage{graphicx}
\usepackage{xcolor}
\usepackage{multirow}
\usepackage[T1]{fontenc}
\usepackage{ae,aecompl}
\usepackage{newtxtext,newtxmath}

\def\msig{$M_{\rm BH}-\sigma$\ }

\def\obj{SDSS J1607+3319}

\title[Opening angle of Dust Torus]
{A practicable estimation of opening angle of dust torus in Type-1.9 AGN with double-peaked broad H$\alpha$}
\author[Zhang]
{Xue-Guang Zhang$^{1}$\thanks{Corresponding author Email: 
	\href{mailto:aexueguang@qq.com}{aexueguang@qq.com}}\\
      $^1$ School of Physical Science and Technology, GuangXi University, No. 100, 
      	Daxue Road, 530004, Nanning, P. R. China}% \\
\pubyear{2022}

\begin{document}

\label{firstpage}

\pagerange{\pageref{firstpage}--\pageref{lastpage}}

\maketitle

\begin{abstract} %%%about 188 words
      In this manuscript, an independent method is proposed to estimate opening angle of dust torus in AGN, 
through unique properties of Type-1.9 AGN with double-peaked broad H$\alpha$ (Type-1.9 DPAGN) coming from 
central accretion disk. Type-1.9 AGN without broad H$\beta$ can be expected by the commonly accepted unified 
model of AGN, considering central BLRs seriously obscured by dust torus with its upper boundary in the line of sight. 
For the unique Type-1.9 DPAGN, accretion disk originations of double-peaked broad H$\alpha$ can be applied 
to determine the inclination angle of the central accretion disk, which is well accepted as substitute of the 
half opening angle of the central dust torus. Then, among low redshift Type-1.9 DPAGN in SDSS, SDSS J1607+3319 
at redshift 0.063 is collected, and the half opening angle of the central dust torus is determined to be 
around 46$\pm$4\degr, after considering disfavoured BBH system to explain the double-peaked broad H$\alpha$ 
through long-term none variabilities and disfavoured local physical conditions to explain disappearance of 
broad H$\beta$ through virial BH mass properties. The results indicate that more detailed studying on dust 
torus of AGN can be appropriately done through Type-1.9 DPAGN in the near future.  
\end{abstract}


\section{Introduction}

	An unified model of Active Galactic Nuclei (AGN) is well known and widely accepted to explain 
different spectroscopic phenomena between Type-1 AGN with optical both broad and narrow emission lines 
and Type-2 AGN with only optical narrow emission lines, after mainly considering obscurations on central 
Broad Line Regions (BLRs) by central dust torus. The unified model has been firstly discussed in 
\citet{an93, um95}, and more recently reviewed and discussed in \citet{nh15, kw21, zh22a}. The Unified 
model has been strongly supported by clearly detected polarized broad emission lines and/or clearly 
detected broad infrared emission lines in some Type-2 AGN \citep{tr03, sg18, mb20}. Moreover, there are 
observational/theoretical evidence to support central dust torus as one fundamental structure in the 
unified model, such as the results in NGC1068\ in \citet{rr98, ma00, gr15} through direct Near-IR images 
and polarimetric images, the resolved dust torus in the Circinus galaxy in \citet{tm07}, the reported 
diversity of dusty torus in AGN in \citet{bm13}, the estimated covering factors of central dust torus 
in local AGN in \citet{eg17}, the determined size of central dust torus in H0507+164\ in \citet{mr18}, 
the well discussed X-ray clumpy torus model in \citet{ou21} etc.. More recent review on dust torus can 
be found in \citet{ar17}.

	Under the framework of the unified model, considering different orientations of central dust 
torus in the line of sight, there is a special kind of AGN, Type-1.9 AGN (firstly discussed in \citet{os81}), 
with broad H$\alpha$ emission lines but no broad H$\beta$ indicating central BLRs seriously obscured by 
dust torus with its upper boundary in the line of sight, besides the Type-1 and Type-2 AGN. Commonly, 
as a transition type, Type-1.9 AGN are considered as the best candidates on studying properties, 
especially properties of spatial structures, of the unified model expected central dust torus. Actually, 
there are some reports on the opening angles (covering factor) of the central dust torus in the literature. 
\citet{at05} have proposed a receding torus model, based on statistically significant correlation between 
the half opening angle of the torus and [O~{\sc iii}] emission-line luminosity, and then followed and 
discussed in \citet{sc05, ar11, mg16, mi19}. \citet{zh18} have reported that the half opening angle of 
the torus declines with increasing accretion rate until the Eddington ratio reaches 0.5, above which 
the trend reverses. \citet{nl16, sr16} have found no evidence for a luminosity dependence of the torus 
covering factor in AGN not to support the receding torus model, similar conclusions can also be found 
in \citet{mc17}. More recent interesting discussions on central obscurations by dust torus can be found in 
\citet{ra22} to support a radiation-regulated unification model in AGN.

	Until now, there are rare reports on the opening angles of the central dust torus in AGN through 
direct spatial resolved images. How to measure/determine the opening angle of the central dust torus 
in an individual AGN is still an interesting challenge. Here, based on unique properties of Type-1.9 
AGN with BLRs being seriously obscured by the central dust torus, an independent method is proposed to 
estimate the opening angle of the central dust torus in a special kind of Type-1.9 AGN, the Type-1.9 
AGN with double-peaked broad H$\alpha$ (Type-1.9 DPAGN). The manuscript is organized as follows. 
Section 2 presents our main hypothesis to estimate the half opening angle of the central dust torus in 
special Type-1.9 DPAGN. Section 3 shows the spectroscopic results of the Type-1.9 DPAGN SDSS 
J160714.40+331909.12 (=\obj) at redshift 0.063. Section 4 gives the main discussions. Section 5 gives 
our final conclusions. And the cosmological parameters have been adopted as
$H_{0}=70{\rm km\cdot s}^{-1}{\rm Mpc}^{-1}$, $\Omega_{\Lambda}=0.7$ and $\Omega_{\rm m}=0.3$.

\section{Main Hypothesis}

	Accretion disk originations have been well accepted to double-peaked broad emission lines, 
as well discussed in \citet{ch89, el95, sn03, sb17}. The inclination angle of the central 
accretion disk can be well estimated through double-peaked broad line emission features. Meanwhile, 
considering the serious obscurations from central dust torus in Type-1.9 DPAGN, the accretion disk 
origination determined inclination angle should be well accepted to trace the half opening angle of 
the central dust torus. Certainly, beside the accretion disk origination, commonly known binary black 
hole (BBH) system can also be applied to explain double-peaked broad emission lines, such as the results 
shown in \citet{sl10}. However, assumed BBH system should lead to optical quasi-periodic oscillations 
(QPOs) with periodicities about hundreds to thousands of days, such as the results shown in \citet{gd15a, 
gm15, zh22}, and will be discussed to disfavor the BBH system in the target in the manuscript. Moreover, 
it should be confirmed that the seriously obscured broad H$\beta$ are not due to local intrinsic physical 
conditions (such as the case in H1320+551 discussed in \citet{bx03}), but due to serious obscurations 
by the central dust torus.

	It is exciting to check whether the method can be applied to estimate the opening angle of 
the central dust torus in Type-1.9 DPAGN, which is the main objective of the manuscript. And the following 
three criteria are accepted to collect targets of the manuscript. First, the targets are Type-1.9 DPAGN, 
with apparent double-peaked broad H$\alpha$ but no apparent broad H$\beta$. Second, there are no signs 
for optical QPOs in the targets, indicating BBH systems not preferred to explain the double-peaked broad 
H$\alpha$. Third, after considering the BH mass properties which will be well discussed in the Section 4, 
serious obscurations by the central dust torus are well accepted to explain the seriously obscured broad 
H$\beta$ in the targets.

	Among the low redshift ($z<0.35$) broad line AGN listed in \citet{sh11} with SPECIAL\_INTEREST\_FLAG=1 
and in \citet{liu19} with flag MULTI\_PEAK=2, there are 561 low redshift DPAGN with reliable broad H$\alpha$ 
emission lines (both reported line width and line luminosity at least five times larger than their reported 
uncertainties). Based on the main hypothesis and corresponding criteria above, Type-1.9 DPAGN SDSS 
J160714.40+331909.12 (=\obj) at redshift 0.063 is collected as the unique target of the manuscript, based 
on two main unique features through properties of its spectroscopic and long-term variabilities, well 
discussed in the next section. On the one hand, among the Type-1.9 DPAGN, the \obj~ has the most apparent 
double-peaked features in broad H$\alpha$. On the other hand, there are no apparent variabilities in \obj, 
which can be well applied to disfavour the BBH system in the \obj, combining its double-peaked features in 
broad H$\alpha$.

\begin{figure}
\centering\includegraphics[width = 8cm,height=7cm]{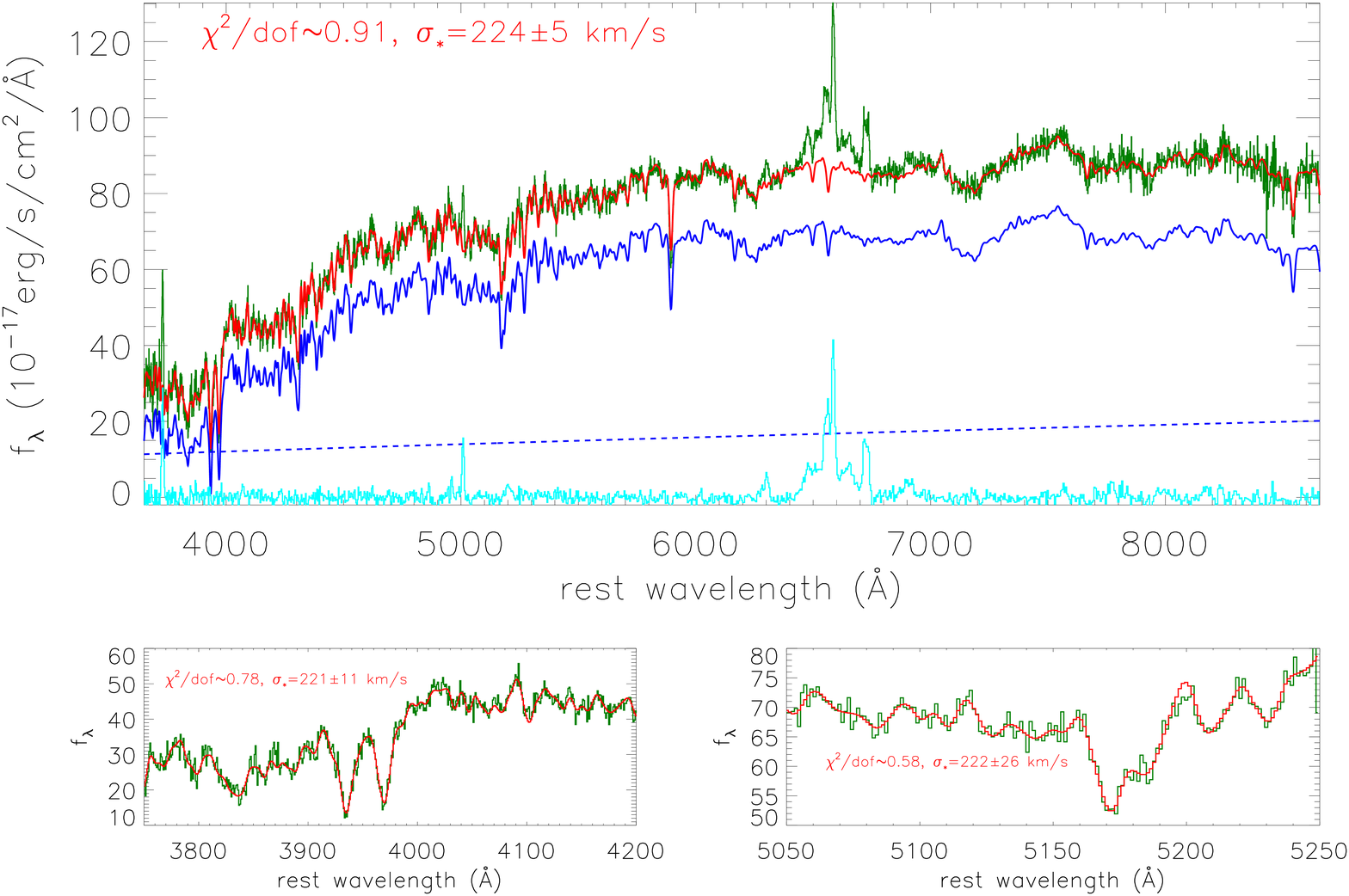}
\caption{Top panel shows the SSP method determined descriptions (solid red line) to the SDSS spectrum 
(solid dark green line) with emission lines being masked out. In top panel, solid blue line and dashed 
blue line show the determined host galaxy contributions and power law AGN continuum emissions, respectively, 
solid cyan line shows the line spectrum calculated by the SDSS spectrum minus the sum of host galaxy 
contributions and AGN continuum emissions. Bottom panels show the best fitting results (solid red line) 
to absorption features (solid dark green line) of Ca~{\sc ii} H+K (left panel), Mg~{\sc i} (right panel). 
In each panel, the determined $\chi^2/dof$ and stellar velocity dispersion are marked in red characters.
}
\label{spec}
\end{figure}

\begin{figure*}
\centering\includegraphics[width = 18cm,height=5cm]{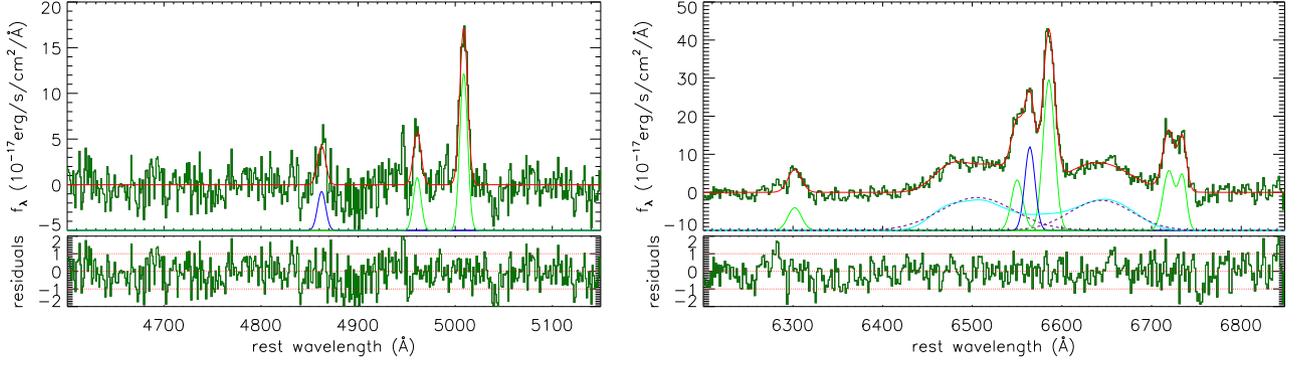}
\caption{Top panels show the best fitting results (solid red line) to the emission lines (solid dark green 
line), and bottom panels show the corresponding residuals. In top left panel, solid blue line shows the 
determined narrow H$\beta$, solid green lines show the determined [O~{\sc iii}] doublet. In top right panel, 
solid blue line shows the determined narrow H$\alpha$, solid cyan line shows the determined double-peaked 
broad H$\alpha$ described by the elliptical accretion disk model, solid green lines show the determined 
[O~{\sc i}], [N~{\sc ii}] and [S~{\sc ii}] doublets, dashed purple lines show the determined broad H$\alpha$ 
described by two broad Gaussian functions. In each bottom panel, horizontal dashed lines show residuals=$0,~\pm1$, 
respectively.
}
\label{line}
\end{figure*}	

\begin{figure*}
\centering\includegraphics[width = 18cm,height=4cm]{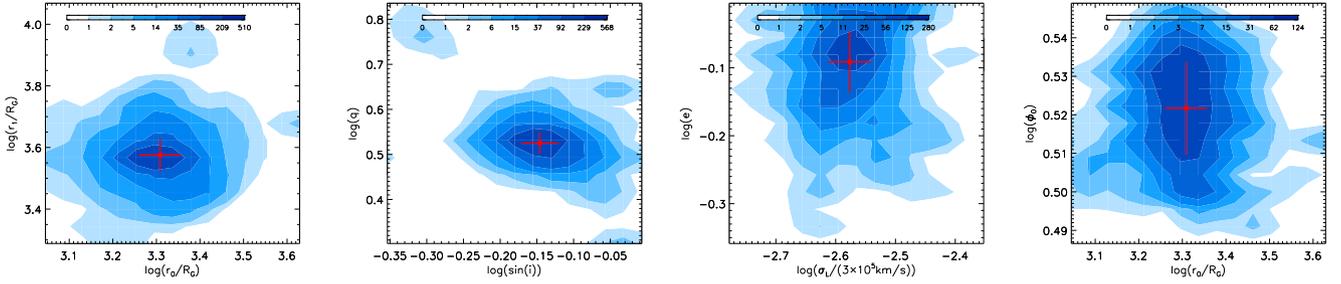}
\caption{MCMC technique determined two-dimensional posterior distributions in contour of the model 
parameters in the elliptical accretion disk model applied to describe the double-peaked broad H$\alpha$. 
In each panel, sold circle plus error bars in red mark the positions of the accepted values and 
corresponding uncertainties of the model parameters. The number densities related to different colors 
are shown in color bar in top region of each panel.}
\label{mcmc}
\end{figure*}

\section{Spectroscopic results of the Type-1.9 DPAGN \obj}

	\obj~ has its SDSS spectrum (plate-mjd-fiberid=1419-53144-0453) with signal-to-noise about 34 
shown in Fig.~\ref{spec}. In order to measure emission lines, the commonly accepted SSP (Simple Stellar 
Population) method is applied to determine host galaxy contributions. More detailed descriptions on the 
SSP method can be found in \citet{bc03, ka03, cm05, cm17}. And the SSP method has been applied in our 
previous papers \citet{zh21a, zh21b, zh21m, zh22a, zh22b}. Here, we show simple descriptions on SSP 
method as follows. The 39 simple stellar population templates from \citet{bc03, ka03} have been exploited, 
combining with a power law component applied to describe intrinsic AGN continuum emissions. When the SSP 
method is applied, optical narrow emission lines are masked out by full width at zero intensity about 
450 ${\rm km/s}$, and the spectrum with wavelength range from 6250 to 6750\AA~ are also masked out 
due to the strongly broad H$\alpha$. Then, through the Levenberg-Marquardt least-squares minimization 
technique, SDSS spectra with emission lines being masked out can be well described by combinations of 
broadened stellar population templates and the power law component. The best descriptions are shown in 
Fig.~\ref{spec} with $\chi^2/dof\sim0.91$ (the summed squared residuals divided by degree of freedom) 
and with determined stellar velocity dispersion (the broadening velocity) about 224$\pm$5~${\rm km/s}$.

	Moreover, in order to determine reliable stellar velocity dispersion, absorption features of 
around Ca~{\sc ii} H+K from 3750 to 4200\AA~ and around Mg~{\sc i} from 5050 to 5250\AA~ are applied 
to re-measure stellar velocity dispersions, through the same SSP method above. The best fitting results 
are shown in bottom panels of Fig.~\ref{spec} with determined stellar velocity dispersions in units of 
${\rm km/s}$ about 222$\pm$11 and 208$\pm$26 through the Ca~{\sc ii} H+K and Mg~{\sc i}, respectively. 
Therefore, in the manuscript, the inverse variance weighted mean stellar velocity dispersion 
$\sigma_\star=$222$\pm$26~${\rm km/s}$ in \obj~ is accepted, which is consistent with the SDSS pipeline 
reported 230~${\rm km/s}$.

	After subtractions of host galaxy contributions and AGN continuum emissions, emission lines in 
the line spectrum can be well measured. Similar as what we have previously done in \citet{zh21a, zh21b, 
zh22a, zh22b, zh22c}, for the emission lines within rest wavelength range from 4600 to 5150\AA, there 
are one broad and one narrow Gaussian functions applied to describe probable broad and apparent narrow 
H$\beta$, two Gaussian functions applied to describe [O~{\sc iii}]$\lambda4959,5007$\AA~ doublet. When 
the functions above are applied, each component has line intensity not smaller than zero, and the 
[O~{\sc iii}] components have the same redshift and the same line width and have flux ratio to be fixed 
to the theoretical value 3. Then, through the Levenberg-Marquardt least-squares minimization technique, 
the best fitting results to the emission lines and the corresponding residuals (line spectrum minus the 
best fitting results and then divided by uncertainties of SDSS spectrum) are shown in left panels of 
Fig.~\ref{line} with $\chi^2/dof\sim0.70$. Based on the fitting results, it is not necessary to consider 
broad Gaussian component in H$\beta$, because the determined line width and line flux (around to zero) 
of the broad Gaussian component are smaller than their corresponding uncertainties, indicating there are 
no apparent broad H$\beta$ in \obj.

	Meanwhile, Gaussian functions can be applied to describe the narrow emission lines within rest 
wavelength range from 6200 to 6850\AA, the [O~{\sc i}], [N~{\sc ii}], [S~{\sc ii}] and narrow H$\alpha$. 
But the commonly accepted elliptical accretion disk model with seven model parameters well discussed in 
\citet{el95} is applied to describe the double-peaked broad H$\alpha$, because the model can be applied 
to explain almost all observational double-peaked broad H$\alpha$ of the \obj. The seven model parameters 
are inner and out boundaries [$r_0$,~$r_1$] in the units of $R_G$ (Schwarzschild radius), inclination 
angle $i$ of disk-like BLRs, eccentricity $e$, orientation angle $\phi_0$ of elliptical rings, local 
broadening velocity $\sigma_L$ in units of ${\rm km/s}$, line emissivity slope $q$ ($f_r~\propto~r^{-q}$). 
Meanwhile, we have also applied the very familiar elliptical accretion disk model in our more recent 
studies on double-peaked lines in \citet{zh21c, zh22a}, and there are no further discussions on the 
elliptical accretion disk model in the manuscript. Then, in order to obtain more reliable uncertainties of 
model parameters in the complicated model functions, rather than the Levenberg-Marquardt least-squares 
Minimization technique, the Maximum Likelihood method combining with the MCMC (Markov Chain Monte Carlo) 
technique \citep{fh13} is applied. The evenly prior distributions of the seven model parameters in the 
elliptical accretion disk model are accepted with the following limitations, $\log(r_0)\in[2,~4]$,
~$\log(r_1)\in[2,~6]$ ($r_1~>~r_0$),~$\log(\sin(i))\in[-3,~0]$,~$\log(q)\in[-1,~1]$,~$\log(\sigma_L)\in[2,~4]$,
~$\log(e)\in[-5,~0]$,~$\log(\phi_0)\in[-5,~\log(2\times\pi)]$. The determined best fitting results and 
corresponding residuals to the emission line around H$\alpha$ are shown in right panels of Fig.~\ref{line} 
with $\chi^2/dof\sim0.48$. The MCMC technique determined posterior distributions of the model parameters 
in the elliptical accretion disk model are shown in Fig.~\ref{mcmc}. And the half width at half maximum 
of each parameter distribution is accepted as uncertainty of the parameter. The determined parameters 
and corresponding uncertainties of each model parameter are listed in Table~1. Moreover, as discussed 
in \citet{zh22a}, clean double-peaked broad line emission features can lead to solely determined model 
parameters in the elliptical accretion disk model. Therefore, there are no further discussions on 
whether is there solely determined model parameter of $\sin(i)$.

\begin{table}
\caption{parameters of the emission line components}
\begin{tabular}{lccc}
\hline\hline
\multicolumn{4}{c}{model parameters of elliptical accretion disk model for broad H$\alpha$} \\
	\multicolumn{4}{c}{$r_0=2035\pm240$,~ $r_1=3766\pm500$,~$\sin(i)=0.71\pm0.04$} \\
	\multicolumn{4}{c}{$q=3.35\pm0.19$,~$e=0.81\pm0.08$,~$\sigma_L=796\pm70{\rm km/s}$,
	~$\phi_0=190\pm6\degr$} \\
\hline\hline
\multicolumn{4}{c}{model parameters of Gaussian emission components} \\
	line & $\lambda_0$ & $\sigma$ & flux \\
\hline
\multirow{2}{*}{broad H$\alpha$} & 6505.6$\pm$1.1 & 41.4$\pm$1.2 & 897$\pm$25 \\
                                 & 6643.9$\pm$1.1 & 34.9$\pm$1.2 & 699$\pm$24 \\
\hline
Narrow H$\alpha$ & 6564.2$\pm$0.5 & 5.6$\pm$0.6 &  311$\pm$54 \\  
\hline
Narrow H$\beta$ & 4862.4$\pm$0.3 & 4.2$\pm$0.4 &  45$\pm$8 \\
\hline
[O~{\sc iii}]$\lambda5007$\AA & 5008.8$\pm$0.3 & 3.9$\pm$0.3 & 172$\pm$10 \\
\hline
[O~{\sc i}]$\lambda6300$\AA & 6301.9$\pm$1.1 & 7.6$\pm$1.2 &  113$\pm$14 \\
\hline
[N~{\sc ii}]$\lambda6583$\AA & 6585.5$\pm$0.2 & 6.5$\pm$0.3 & 642$\pm$55 \\
\hline
[S~{\sc ii}]$\lambda6716$\AA & 6719.2$\pm$0.9 & 6.6$\pm$0.9 &  260$\pm$33 \\
\hline
[S~{\sc ii}]$\lambda6731$\AA & 6734.5$\pm$0.8 & 4.6$\pm$0.7 & 157$\pm$29 \\
\hline\hline
\end{tabular}\\
Notice: For the Gaussian emission components, the first column shows which line is measured, the 
Second, third, fourth columns show the measured line parameters: the center wavelength $\lambda_0$ 
in unit of \AA, the line width (second moment) $\sigma$ in unit of \AA~ and the line flux in unit 
of ${\rm 10^{-17}~erg/s/cm^2}$. 
\end{table}

\section{Main Discussions}

        In the section, two points are mainly considered. First, it is necessary to determine that 
the accretion disk origination is favoured to explain the double-peaked broad H$\alpha$ in \obj, rather 
than a BBH system. Second, it is necessary to determine that the large broad Balmer decrement (flux 
ratio of broad H$\alpha$ to broad H$\beta$) is due to serious obscurations, rather than due to local 
physical conditions, because that BLRs modeled with relatively low optical depths and low ionization 
parameters can reproduce large broad Balmer decrements, as well discussed in \citet{kk81, cp81, gr90} 
without considering serious obscurations and see the unobscured central regions in a Type-1.9 AGN in 
\citet{bx03}.

\begin{figure}
\centering\includegraphics[width = 8cm,height=5cm]{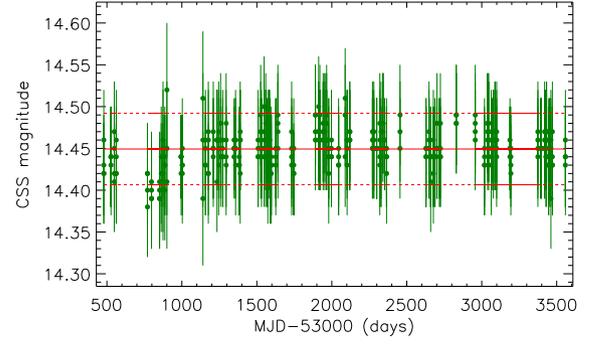}
\caption{CSS V-band light curve of \obj. Horizontal solid and dashed red lines show the mean value and 
corresponding 2RMS scatters of the light curve.}
\label{lmc}
\end{figure}

	For the first point on BBH system, the following discussions are given. The double-peaked broad 
H$\alpha$ can also be well described by two broad Gaussian functions shown as dashed purple lines in top 
right panel of Fig.~\ref{line} with model parameters listed in Table~1. Under the assumption of BBH system 
in \obj, considering the strong linear correlation between broad H$\alpha$ luminosity and continuum 
luminosity as discussed in \citet{gh05}, there are totally equal (ratio about 897:699 from emission fluxes 
of the two broad Gaussian components) continuum luminosities related to central two BH accreting systems, 
indicating there should be strong variabilities with QPOs due to orbital rotating effects. However, there 
are none variabilities in the collected 8.4years-long CSS (Catalina Sky Survey, \citet{dd09}) V-band light 
curve shown in Fig.~\ref{lmc} with almost all data points lying within 2RMS scatter ranges. Therefor, 
rather than the BBH system, the elliptical accretion disk model is preferred to explain the double-peaked 
broad H$\alpha$ in \obj.

\begin{figure}
\centering\includegraphics[width = 8cm,height=5cm]{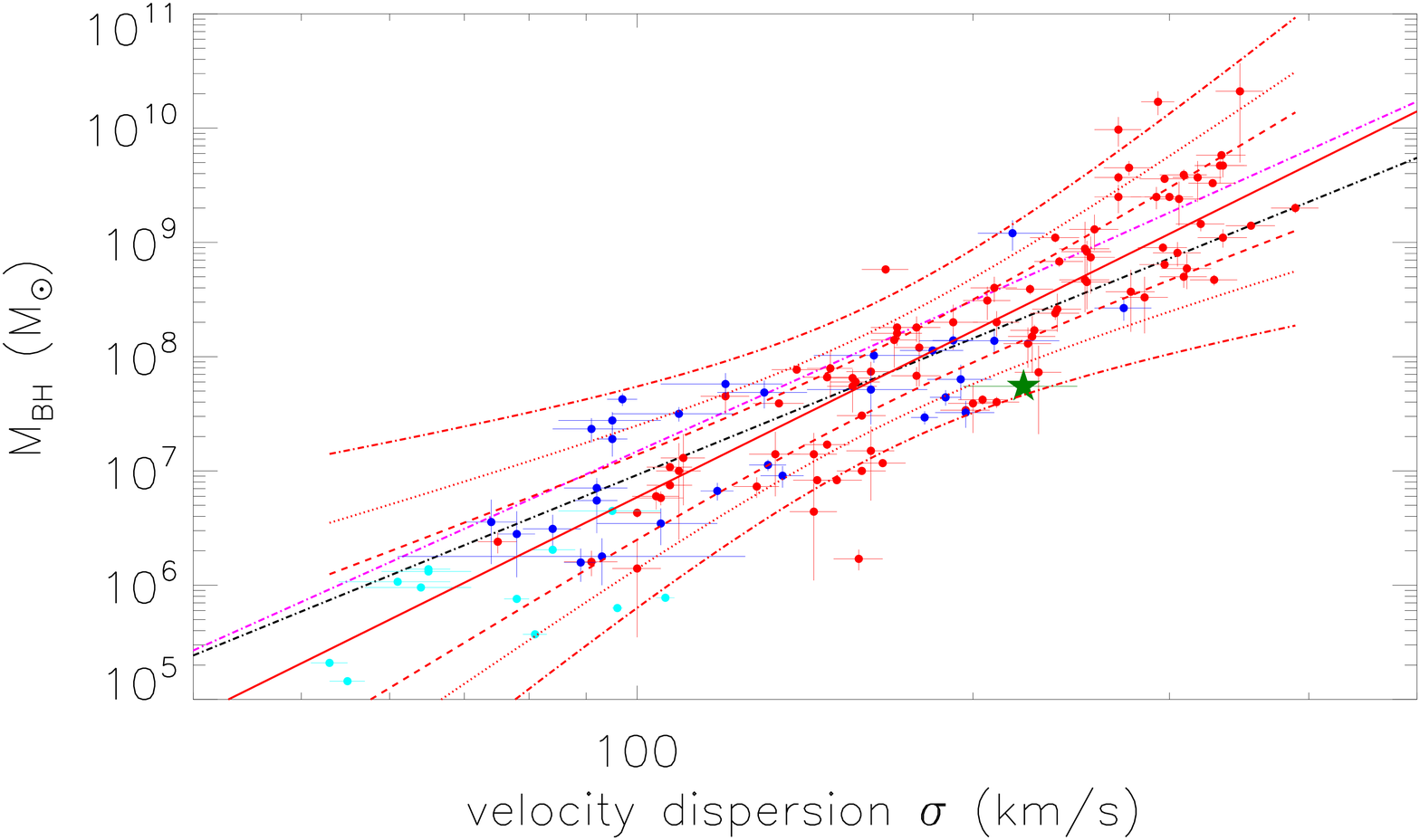}
\caption{On the correlation between stellar velocity dispersion measured through absorption features and 
virial BH mass of \obj. Solid five-point-star in dark green shows the virial BH mass of \obj~ determined 
by properties of observed broad H$\alpha$. Dot-dashed lines in magenta and in black represent the \msig 
relations through the quiescent galaxies in \citet{kh13} and through the RM AGNs in \citet{wy15}, respectively. 
Solid circles in red, in blue and in pink show the values for the 89 quiescent galaxies in \citet{sg15}, 
the 29 RM AGNs in \citet{wy15} and the 12 TDEs in \citet{zl21}, respectively. Thick solid red line shows 
the best fitting results to all the objects, and thick dashed, dotted and dot-dashed red lines show 
corresponding $3\sigma$, $4\sigma$ and $5\sigma$ confidence bands to the best fitting results.}
\label{msig}
\end{figure}

%%%line profile of broad Halpha, wavelength range from 6350 to 6800
	For the second point, properties of virial BH mass are mainly discussed. Based on accepted 
virialization assumptions to properties of observed broad H$\alpha$ as discussed in \citet{ve02, pe04, gh05, 
sh11, mt22}, virial BH mass can be estimated by 
\begin{equation}
\begin{split}
M_{BH}&=15.6\times10^6(\frac{L_{H\alpha}}{\rm 10^{42}erg/s})^{0.55}
	(\frac{\sigma_{H\alpha}}{\rm 1000km/s})^{2.06}{\rm M_\odot} \\
	&=(5.5\pm0.6)\times10^7{\rm M_\odot}
\end{split}
\end{equation}
with $L_{H\alpha}=(1.39\pm0.05)\times10^{41}{\rm erg/s}$ as line luminosity of observed broad H$\alpha$ 
and $\sigma_{H\alpha}=(3100\pm110){\rm km/s}$ as second moment of observed broad H$\alpha$, after considering 
more recent empirical R-L relation to estimate BLRs sizes in \citet{bd13}. Uncertainty of virial BH mass 
is determined by uncertainties of the $L_{H\alpha}$ and $\sigma_{H\alpha}$. If large broad Balmer decrement 
was due to local physical conditions, the estimated virial BH mass should be simply consistent with the 
\msig relation \citep{fm00, ge00, kh13, bb17, bt21} expected value, otherwise, there should be smaller 
virial BH mass. Then, Fig.~\ref{msig} shows virial BH mass properties of \obj~ in the \msig space. In order 
to show clearer results, the 89 quiescent galaxies from \citet{sg15} and the 29 reverberation mapped (RM) 
AGN from \citet{wy15} and the 12 tidal disruption events (TDEs) from \citet{zl21} are considered to draw 
the linear correlation between stellar velocity dispersion and BH mass
\begin{equation}
\log(\frac{M_{BH}}{\rm M_\odot})~=~(-2.89\pm0.49)~+~
	(4.83\pm0.22)\times\log(\frac{\sigma_\star}{\rm km/s})
\end{equation}
through the Least Trimmed Squares robust technique \citep{cm13}. And then the $3\sigma$, $4\sigma$ and 
$5\sigma$ confidence bands to the linear correlation are determined and shown in Fig.~\ref{msig}. Therefore, 
the estimated viral BH mass of \obj~ is lower than \msig expected value with confidence level higher than 
$4\sigma$. Therefore, locate physical conditions are disfavored to explain the large broad Balmer decrement 
in \obj.

	Based on the double-peaked broad H$\alpha$ in the Type-1.9 DPAGN \obj, half opening angle of 
central dust torus is well estimated as (46$\pm$4)\degr~($\sin(i)\sim0.71\pm0.04$), roughly consistent 
with statistical mean value in \citet{zh18}. Therefore, it is interesting to study properties of opening 
angles of dust torus through Type-1.9 DPAGN in the near future, after many efforts to disfavour BBH 
systems to explain their double-peaked broad H$\alpha$ and to disfavour local physical conditions to 
explain disappearance of broad H$\beta$.

	Before ending of the manuscript, an additional point is noted. Before giving clear physical 
information of materials in the central dust torus, it is hard to confirm that the accretion disk 
origination determined inclination angle is completely consistent with the half opening angle of the 
central dust torus in Type-1.9 DPAGN. If material densities in regions around upper boundary of the 
central dust torus were too low to lead the broad H$\beta$ being totally obscured, the determined 
inclination angle should be lower than the intrinsic half opening angle of the central dust torus. 
Moreover, it is not clear whether are there different radial dependent material densities in the 
direction perpendicular to the equatorial plane related to central AGN activities, which should also 
have effects on the consistency between the accretion disk origination determined inclination angle 
and the half opening angle of the central dust torus in AGN with different central AGN activities. 
In the near future, through studying a sample of Type-1.9 DPAGN as one of our ongoing projects, clearer 
clues and detailed discussions will be given on the consistency between the inclination angle and the 
half opening angle of the central dust torus.

\section{Conclusions}

	An independent method is proposed to estimate the opening angle of the central dust torus in 
Type-1.9 DPAGN through unique double-peaked features of broad H$\alpha$, accepted the assumptions of 
obscurations of the central dust torus on BLRs leading to disappearance of broad H$\beta$ and of the 
double-peaked broad H$\alpha$ with accretion disk originations. Then, among the reported DPAGN, the 
\obj~ is collected due to its apparent broad double-peaked broad H$\alpha$ but no broad H$\beta$. 
Moreover, long-term optical variabilities can be applied to disfavour the BBH system in \obj~ to explain 
the double-peaked broad H$\alpha$. And properties of virial BH mass can be applied to determine that 
local physical conditions are not favoured to explain the large broad Balmer decrement in \obj. Then, 
based on the well applied elliptical accretion disk model applied to describe the double-peaked broad 
H$\alpha$ in \obj, the half opening angle of the central dust torus can be well estimated as 
(46$\pm$4)\degr\ in \obj. The results in the manuscript strongly indicate that the proposed independent 
method is practicable, and can be applied to study detailed properties of the opening angles of the 
central dust torus through a sample of Type-1.9 DPAGN, which will be studied in the near future.

\section*{Acknowledgements}
Zhang gratefully acknowledges the anonymous referee for giving us constructive comments and suggestions 
to greatly improve our paper. Zhang gratefully acknowledges the kind funding support NSFC-12173020. 
This research has made use of the data from the SDSS (\url{https://www.sdss.org/}) funded 
by the Alfred P. Sloan Foundation, the Participating Institutions, the National Science Foundation and 
the U.S. Department of Energy Office of Science, and use of the data from CSS 
\url{http://nesssi.cacr.caltech.edu/DataRelease/}. The research has made use of the MPFIT package 
\url{https://pages.physics.wisc.edu/~craigm/idl/cmpfit.html}, and of the LTS\_LINEFIT package 
\url{https://www-astro.physics.ox.ac.uk/~cappellari/software/}, and of the emcee package 
\url{https://pypi.org/project/emcee/}.

\section*{Data Availability}
The data underlying this article will be shared on request to the corresponding author
(\href{mailto:aexueguang@qq.com}{aexueguang@qq.com}).

\label{lastpage}

\begin{thebibliography}{   }
\bibitem[\protect\citeauthoryear{Antonucci}{1993}]{an93}
Antonucci, R., 1993, ARA\&A, 31, 473	
\bibitem[\protect\citeauthoryear{Almeida \& Ricci}{2017}]{ar17}
Almeida, C. R., Ricci, C., 2017, Nat Astron, 1, 679
\bibitem[\protect\citeauthoryear{Alonso-Herrero et al.}{2011}]{ar11}
Alonso-Herrero, A.; Ramos Almeida, C.; Mason, R., et al., 2011, ApJ, 736, 82
\bibitem[\protect\citeauthoryear{Arshakian}{2005}]{at05}
Arshakian, T. G., 2005, A\&A, 436, 817
\bibitem[\protect\citeauthoryear{Batiste et al.}{2017}]{bb17}
Batiste, M.; Bentz, M. C.; Raimundo, S. I.; Vestergaard, M.; Onken, C. A., 2017, ApJL, 838, 10
\bibitem[\protect\citeauthoryear{Bennert et al.}{2021}]{bt21}
Bennert, V. N.; Treu, T.; Ding, X.; et al., 2021, ApJ, 921, 36
%\bibitem[\protect\citeauthoryear{Balokovic et al.}{2018}]{bb18}UM
%Balokovic, M.; Brightman, M.; Harrison, F. A.; et al., 2018, ApJ, 854, 42
\bibitem[\protect\citeauthoryear{Barcons et al.}{2003}]{bx03}
Barcons, X.; Carrera, F. J.; Ceballos, M. T., 2003, MNRAS, 339, 757
\bibitem[\protect\citeauthoryear{Bentz et al.}{2013}]{bd13}
Bentz, M. C.; Denney, K. D.; Grier, C. J, et al., 2013, ApJ, 767, 149
%\bibitem[\protect\citeauthoryear{Bornancini \& Garcia Lambas}{2020}]{bg20}
%Bornancini, C.; Garcia Lambas, D., 2020, MNRAS, 494, 1189
\bibitem[\protect\citeauthoryear{Bruzual \& Charlot}{2003}]{bc03}
Bruzual, G.; Charlot, S. 2003, MNRAS, 344, 1000
\bibitem[\protect\citeauthoryear{Burtscher}{2013}]{bm13}
Burtscher, L.; Meisenheimer, K.; Tristram, K. R. W., et al., 2013, A\&A, 558, 149
\bibitem[\protect\citeauthoryear{Canfield \& Puetter}{1981}]{cp81}
Canfield, R. C.; Puetter, R. C., 1981, ApJ, 243, 390
\bibitem[\protect\citeauthoryear{Cappellari et al.}{2013}]{cm13}
Cappellari, M.; Scott, N.; Alatalo, K., et al., 2013, MNRAS, 432, 1709
\bibitem[\protect\citeauthoryear{Cappellari}{2017}]{cm17}
Cappellari, M., 2017, MNRAS, 466, 798
\bibitem[\protect\citeauthoryear{Chen \& Halpern}{1989}]{ch89}
Chen, K. Y., \& Halpern, J. P., 1989, ApJ, 344, 115
\bibitem[\protect\citeauthoryear{Cid Fernandes et al.}{2005}]{cm05}
Cid Fernandes, R.; Mateus, A.; Sodre, L.; Stasinska, G.; Gomes, J. M., 2005, MNRAS, 358, 363
\bibitem[\protect\citeauthoryear{Drake et al.}{2009}]{dd09}
Drake, A. J.; Djorgovski, S. G.; Mahabal, A.; et al., 2009, ApJ, 696, 870
\bibitem[\protect\citeauthoryear{Eracleous et al.}{1995}]{el95}
Eracleous, M., Livio, M., Halpern, J. P., Storchi-Bergmann, T.,1995, ApJ, 438, 610
\bibitem[\protect\citeauthoryear{Ezhikode et al.}{2017}]{eg17}
Ezhikode, S. H.; Gandhi, P.; Done, C.; Ward, M.; Dewangan, G. C.; Misra, R.; Philip, N. S, 2017, MNRAS, 472, 3492
\bibitem[\protect\citeauthoryear{Ferrarese \& Merritt}{2000}]{fm00}
Ferrarese, F.; Merritt, D., 2000, ApJL, 539, 9
%\bibitem[\protect\citeauthoryear{Franceschini et al.}{2002}]{fb02}
%Franceschini, A.; Braito, V.; Fadda, D., 2002, MNRAS Letter, 335, 51
\bibitem[\protect\citeauthoryear{Foreman-Mackey et al.}{2013}]{fh13}
Foreman-Mackey, D.; Hogg, D. W.; Lang, D.; Goodman, J., 2013, PASP, 125, 306
%\bibitem[\protect\citeauthoryear{Fritz et al.}{2006}]{ff06}  UM
%Fritz, J.; Franceschini, A.; Hatziminaoglou, E., 2006, MNRAS, 366, 767
\bibitem[\protect\citeauthoryear{Gebhardt et al.}{2000}]{ge00}
Gebhardt, K.; Bender, R.; Bower, G, et al., 2000, ApJL, 539, 13
\bibitem[\protect\citeauthoryear{Graham et al.}{2015a}]{gd15a}
Graham, M. J.; Djorgovski, S. G.; Stern, D., et al., 2015a, Natur, 518, 74
\bibitem[\protect\citeauthoryear{Graham et al.}{2015b}]{gm15}
Graham, M. J., Djorgovski, S. G., Stern, D., et al., 2015b, MNRAS, 453, 1562
\bibitem[\protect\citeauthoryear{Gratadour et al.}{2015}]{gr15}
Gratadour, D.; Rouan, D.; Grosset, L.; Boccaletti, A.; Clenet, Y., 2015, A\&A, 581, 8
\bibitem[\protect\citeauthoryear{Greene \& Ho}{2005}]{gh05}
Greene, J. E.; Ho, L. C., 2005, ApJ, 630, 122
\bibitem[\protect\citeauthoryear{Goodrich}{1990}]{gr90}
Goodrich, R. W., 1990, ApJ, 355, 88
%\bibitem[\protect\citeauthoryear{Graham et al.}{2011}]{gr11}
%Graham, A. W.; Onken, C. A.; Athanassoula, E.; Combes, F. 2011, MNRAS, 412, 2211
%\bibitem[\protect\citeauthoryear{Ho \& Kim}{2014}]{hk14}
%Ho, L. C.; Kim, M.-J., 2014, ApJ, 789, 17
%\bibitem[\protect\citeauthoryear{Hernandez-Garcia et al.}{2017}]{hm17}
%Hernandez-Garcia, L.; Masegosa, J.; Gonzalez-Martin, O.; Marquez, I.; Guainazzi, M.; Panessa, F., 2017, A\&A, 602, 65
\bibitem[\protect\citeauthoryear{Kauffmann et al.}{2003}]{ka03}
Kauffmann, G.; Heckman, T. M.; Tremonti, C., et al. 2003, MNRAS, 346, 1055
\bibitem[\protect\citeauthoryear{Kwan \& Krolik}{1981}]{kk81}
Kwan, J.; Krolik, J. H., 1981, ApJ, 250, 478
\bibitem[\protect\citeauthoryear{Kormendy \& Ho}{2013}]{kh13}
Kormendy, J.; Ho, L. C., 2013, ARA\&A, 51, 511
\bibitem[\protect\citeauthoryear{Kuraszkiewicz et al.}{2021}]{kw21}
Kuraszkiewicz, J.; Wilkes, B. J.; Atanas, A.; et al., 2021, ApJ, 913, 134
\bibitem[\protect\citeauthoryear{Liu et al.}{2019}]{liu19}
Liu, H.; Liu, W.; Dong, X.; Zhou, H.; Wang, T.; Lu, H.; Yuan, W., 2019, ApJS, 243, 21
\bibitem[\protect\citeauthoryear{Mandal et al.}{2018}]{mr18}
Mandal, A. K.; Rakshit, S.; Kurian, K. S., et al., 2018, MNRAS, 475, 5330
\bibitem[\protect\citeauthoryear{Marco \& Alloin}{2000}]{ma00}
Marco, O.; Alloin, D., 2000, A\&A, 353, 465
\bibitem[\protect\citeauthoryear{Marin et al.}{2016}]{mg16}
Marin, F.; Goosmann, R. W.; Petrucci, P. O., 2016, A\&A, 519, 23
\bibitem[\protect\citeauthoryear{Mateos et al.}{2017}]{mc17}
Mateos, S.; Carrera, F. J.; Barcons, X., et al., 2017, ApJL, 841, 18
\bibitem[\protect\citeauthoryear{Matt et al.}{2019}]{mi19}
Matt, G.; Iwasawa, K., 2019, MNRAS, 482, 151
%\bibitem[\protect\citeauthoryear{McConnell \& Ma}{2013}]{mm13}
%McConnell, N. J.; Ma, C. P., 2013, ApJ, 764, 184
\bibitem[\protect\citeauthoryear{Mejia-Restrepo et al.}{2022}]{mt22}
Mejia-Restrepo, J. E.; Trakhtenbrot, B.; Koss, M. J., et al., 2022, ApJS, 261, 5
\bibitem[\protect\citeauthoryear{Moran et al.}{2020}]{mb20}
Moran, E. C.; Barth, A. J.; Kay, L. E.; Filippenko, A. V., 2020, ApJL, 540, 73
\bibitem[\protect\citeauthoryear{Netzer}{2015}]{nh15}
Netzer, H., 2015, ARA\&A, 53, 365
\bibitem[\protect\citeauthoryear{Netzer et al.}{2016}]{nl16}
Netzer, H.; Lani, C.; Nordon, R.; Trakhtenbrot, B.; Lira, P.; Shemmer, O., 2016, ApJ, 819, 123
\bibitem[\protect\citeauthoryear{Ogawa et al.}{2021}]{ou21}
Ogawa, S.; Ueda, Y.; Tanimoto, A.; Yamada, S., 2021, ApJ, 906, 84
%\bibitem[\protect\citeauthoryear{Oh et al.}{2015}]{oy15}
%Oh, K.; Yi, S. K.; Schawinski, K.; Koss, M.; Trakhtenbrot, B.; Soto, K., 2015, ApJS, 219, 1
%\bibitem[\protect\citeauthoryear{Onken et al.}{2004}]{On04}
%Onken, C. A.; Ferrarese, L.; Merritt, D.; Peterson, B. M.; Pogge, R. W.;
%	        Vestergaard, M.; Wandel, A., 2004, ApJ, 615, 645
\bibitem[\protect\citeauthoryear{Osterbrock}{1981}]{os81}
Osterbrock, D. E., 1981, ApJ, 249, 462
%\bibitem[\protect\citeauthoryear{Osterbrock \& Mathews}{1986}]{om86}
%Osterbrock, D. E.; Mathews, W. G., 1986, ARA\&A, 24, 171
%\bibitem[\protect\citeauthoryear{Park et al.}{2012}]{pa12}
%Park, D.; Kelly, B. C.; Woo, J.-H.; Treu, T., 2012, ApJS, 203, 6
\bibitem[\protect\citeauthoryear{Peterson et al.}{2004}]{pe04}
Peterson, B. M.; Ferrarese, L.; Gilbert, K. M., et al., 2004, ApJ, 613, 682
%\bibitem[\protect\citeauthoryear{Ramos Almeida et al.}{2011}]{rl11} um
%Ramos Almeida, C.; Levenson, N. A.; Alonso-Herrero, A., et al., 2011, ApJ, 731, 92
\bibitem[\protect\citeauthoryear{Ricci et al.}{2022}]{ra22}
Ricci, C.; Ananna, T. T.; Temple, M. J.; et al., 2022, ApJ, 938, 67
\bibitem[\protect\citeauthoryear{Rouan et al.}{1998}]{rr98}
Rouan, D.; Rigaut, F.; Alloin, D.; Doyon, R.; Lai, O.; Crampton, D.; Gendron, E.; Arsenault, R., 1998, A\&A, 339, 687
\bibitem[\protect\citeauthoryear{Savic et al.}{2018}]{sg18}
Savic, D.; Goosmann, R.; Popovic, L. C.; Marin, F.; Afanasiev, V. L., 2018, A\&A, 614, 120
\bibitem[\protect\citeauthoryear{Savorgnan \& Graham}{2015}]{sg15}
Savorgnan, G. A. D.; Graham, A. W., 2015, MNRAS, 446, 2330
\bibitem[\protect\citeauthoryear{Shen \& Loeb}{2010}]{sl10}
Shen, Y.; Loeb, A., 2010, ApJ, 725, 249
\bibitem[\protect\citeauthoryear{Shen et al.}{2011}]{sh11}
Shen, Y.; Richards, G. T.; Strauss, M. A.; et al., 2011, ApJS, 194, 45
\bibitem[\protect\citeauthoryear{Simpson}{2005}]{sc05}
Simpson, C., 2005, MNRAS, 360, 565
\bibitem[\protect\citeauthoryear{Stalevski et al.}{2016}]{sr16}
Stalevski, M., Ricci, C., Ueda, Y.,  Lira, P., Fritz, J., Baes, M., 2016, MNRAS, 458, 2288
\bibitem[\protect\citeauthoryear{Storchi-Bergmann et al.}{2003}]{sn03}
Storchi-Bergmann, T., et al., 2003, ApJ, 489, 8
\bibitem[\protect\citeauthoryear{Storchi-Bergmann et al.}{2017}]{sb17}
Storchi-Bergmann, T., et al., 2017, ApJ, 835, 236
%\bibitem[\protect\citeauthoryear{Strateva et al.}{2003}]{st03}
%	Strateva, I. V., et al., 2003, AJ, 126, 1720
%\bibitem[\protect\citeauthoryear{Sulentic et al.}{2000}]{sm00}
%Sulentic, J. W.; Marziani, P.; Dultzin-Hacyan, D., 2000, ARA\&A, 38, 521
\bibitem[\protect\citeauthoryear{Tristram et al.}{2007}]{tm07}
Tristram, K. R. W.; Meisenheimer, K.; Jaffe, W., et al., 2007, A\&A, 474, 837
\bibitem[\protect\citeauthoryear{Tran}{2003}]{tr03}
Tran, H. D., 2003, ApJ, 583, 632
\bibitem[\protect\citeauthoryear{Urry \& Padovani}{1995}]{um95}
Urry, C. M.; Padovani, P., 1995, PASP, 107, 803
\bibitem[\protect\citeauthoryear{Vestergaard}{2002}]{ve02}
Vestergaard, M., 2002, ApJ, 571, 733
%\bibitem[\protect\citeauthoryear{Villarroel \& Korn}{2014}]{vk14}
%Villarroel, B.; Korn, A. J., 2014, Nature Physics, 10, 417
%\bibitem[\protect\citeauthoryear{Woo et al.}{2010}]{wt10}
%Woo, J.-H.; Treu, T.; Barth, A. J.; et al., 2010, ApJ, 716, 269
%\bibitem[\protect\citeauthoryear{Woo et al.}{2013}]{ws13}
%Woo, J.-H., Schulze, A.; Park, D.; Kang, W.; Kim, S. C.; Riechers, D. A., 2013, ApJ, 772, 49
\bibitem[\protect\citeauthoryear{Woo et al.}{2015}]{wy15}
Woo, J.; Yoon, Y.; Park, S.; Park, D.; Kim, S. C., 2015, ApJ, 801, 38
\bibitem[\protect\citeauthoryear{Zhang}{2021d}]{zh21m}
Zhang, X. G., 2021d, MNRAS, 502, 2508
\bibitem[\protect\citeauthoryear{Zhang}{2021a}]{zh21a}
Zhang, X. G., 2021a, ApJ, 909, 16, ArXiv:2101.02465  %%%shifted O3
\bibitem[\protect\citeauthoryear{Zhang}{2021b}]{zh21b}
Zhang, X. G., 2021b, ApJ, 919, 13, ArXiv:2107.09214  %%%THE BLUEST CHANGING-LOOK agn
\bibitem[\protect\citeauthoryear{Zhang}{2021c}]{zh21c}
Zhang, X. G., 2021c, MNRAS Letter, 500, 57
\bibitem[\protect\citeauthoryear{Zhang}{2022a}]{zh22a}
Zhang, X. G., 2022a, ApJS, 260, 31   %%NLR sizes of DBP
\bibitem[\protect\citeauthoryear{Zhang}{2022b}]{zh22b}
Zhang, X. G., 2022b, ApJS, 261, 23   %%%stellar velocity dispersions
\bibitem[\protect\citeauthoryear{Zhang}{2022c}]{zh22c}
Zhang, X. G., 2022c, ApJ, 937, 105, ArXiv:2209.02164  %%%True Type-2 QSO
\bibitem[\protect\citeauthoryear{Zhang}{2022}]{zh22}
Zhang, X. G., 2022, MNRAS accepted, Arxiv:2202.11995
\bibitem[\protect\citeauthoryear{Zhou et al.}{2021}]{zl21}
Zhou, Z. Q.; Liu, F. K.; Komossa, S., et al., 2021, ApJ, 907, 77
\bibitem[\protect\citeauthoryear{Zhuang et al.}{2018}]{zh18}
Zhuang, M.; Ho, L. C.; Shangguan, J., 2018, ApJ, 862, 118
%\bibitem[\protect\citeauthoryear{Zou et al.}{2019}]{zy19}
%Zou, F.; Yang, G.; Brandt, W. N.; Xue, Y., 2019, ApJ, 878, 11
\end{thebibliography}
\end{document}